\documentclass[aps,prl,twocolumn,showpacs]{revtex4}
\usepackage{graphicx}
\usepackage{physics}
\usepackage{color}

\def\dvarphi{\frac{\partial}{\partial\varphi}}
\def\d2varphi{\frac{\partial^2}{\partial\varphi^2}}
\def\ce{\mbox{ce}}
\def\se{\mbox{se}}
\def\dn{\mbox{dn}}
\def\cn{\mbox{cn}}

\begin{document}

\title{Zitterbewegung and symmetry switching in  Klein's four-group}
\author{L. Chotorlishvili$^1$, P. Zi\c{e}ba$^2$ I. Tralle$^2$, A. Ugulava$^3$}
\address{$^1$Institut f\"ur Physik, Martin-Luther Universit\"at Halle-Wittenberg, D-06120 Halle/Saale, Germany \\
$^2$  Faculty of Mathematics and Natural Sciences, University of Rzeszow,   Pigonia str. 1, 35-310 Rzeszow, Poland \\
$^3$  Faculty of Mathematics and Natural Sciences, Tbilisi State University, Chavchavadze av.3, 0128 Tbilisi, Georgia}

\begin{abstract}
Zitterbewegung is the exotic phenomenon associated either with the relativistic electron-positron rapid oscillation
or to the electron-hole transitions in the narrow gap semiconductors. In the present work, we enlarge concept of
Zitterbewegung and show that the trembling motion may occur due to the dramatic changes in the symmetry of the system.
In particular, we exploit a paradigmatic model of quantum chaos, quantum mathematical pendulum (universal Hamiltonian). The symmetry group of this system
is the Klein's four-group that possess three invariant subgroups. The energy spectrum of the system parametrically depends
on the height of the potential barrier, and contains degenerate and non-degenerate areas, corresponding to the different symmetry subgroups.
Change in the height of the potential barrier switches the symmetry subgroup and leads to the trembling motion. We analyzed mean square
fluctuations of the velocity operator and observed that trembling enhances for the highly excited states. We  observed the link between the phenomena of trembling motion
and uncertainty relations of noncommutative operators of the system.

\end{abstract}
\maketitle

\section{Introduction}

Zitterbewegung (ZB) is the trembling motion, phenomenon discovered by Schr\"odinger \cite{Schrodinger}. For a long time, ZB was associated
solely with Dirac equation. Due to the purely relativistic nature, oscillation frequency of the relativistic ZB  $2m_{e}c^{2}/\hbar\approx 10^{7}$
THz is far beyond the experimentally detectable
frequencies. Therefore interest to the ZB has cooled down right after its discovery. Nevertheless last decade,
we witness renewed interest to this phenomena. \\

The reason for the renewed interest of ZB is the application of Dirac equation in the non-relativistic condensed matter physics.
In particular in the systems with a spin-orbit coupling (graphene, 2D electron gas, topological insulators)
light velocity is replaced by the velocity of electrons at the Fermi surface. Naturally, this lowers the threshold
frequency of the ZB towards the experimentally detectable scale. The relativistic ZB concerns the electron-positron
rapid transition oscillations, while in the narrow-gap semiconductors one could talk about oscillations due to the
mixing of the conductance and hole band states \cite{zawadzki20111}. ZB of electronic wave packets has been studied
in the semiconductor quantum wells \cite{Schliemann} and in the optical traps \cite{Solano}. \\

Fascinating relativistic effect of condensed matter physics is the Klein paradox in graphene \cite{Katsnelson}.
Due to the matching of the electron and positron wave functions across the barrier, the transmission probability is large even
for high barriers. This effect cannot occur in the non-relativistic case because of exponential decay of the transmission probability with the barrier height.
The Hamiltonian of a single ion trapped in the Paul
trap bares a striking resemblance to the Dirac Hamiltonian \cite{Blatt}. This allows experimental observation of ZB in the cold atom physics
\cite{Engels,LeBlanc}.
Note that in case of the "nonrelativistic" ZB particle performing trembling motion
not necessary is a free particle but can be a particle trapped in the external potential \cite{zawadzki2005,zawadzki2007}.
The role of the truncated Coulomb potential for the 1D Dirac materials has been studied recently \cite{Portnoi}.\\

In the broader sense, ZB may occur in an arbitrary system characterized by anomalous velocity relevant to
the case when momentum operator commutes with Hamiltonian but velocity operator does not commute. In particular the Hamiltonian of the relativistic ZB
reads:
\begin{equation}
\begin{aligned}
\hat{H}=c(\alpha \hat{p})+\gamma_{0}mc^{2},~~~~ \frac{d\hat{p}}{dt}=0,~~~~\frac{d\hat{r}}{dt}=c\hat{\alpha}, \\
\end{aligned}
\end{equation}

\begin{equation}
\begin{aligned}
\alpha &=
 \begin{pmatrix}
  0 & \hat{\sigma^{0}} \\
  \hat{\sigma^{0}} & 0 \\
 \end{pmatrix},
  &
 \gamma &=
 \begin{pmatrix}
  0 & -i\hat{\sigma^{0}} \\
  i\hat{\sigma^{0}} & 0 \\
 \end{pmatrix},
\end{aligned}
\end{equation}
leading to the time dependent linear and trembling  terms
\begin{equation}
x(t)=c^{2}\hat{p_{x}}\hat{H}^{-1}t-\frac{c\hbar^{2}}{4}\dot{\alpha_{1}}(0)\exp\bigg(-\frac{2i\hat{H}t}{\hbar}\bigg)\hat{H}^{-2}.
\end{equation}
Here $\hat{\sigma^{0}}$ is the vector of Pauli matrixes and $\hat{H}$ is the free particle Dirac's Hamiltonian. Thus in general,
motion may have a trembling character if the following commutator is nonzero

\begin{equation}
\frac{d\vec{r}}{dt}=(1/\imath \hbar)\big[\vec{r},\hat{H}\big].
\end{equation}

Typically, the phase space of nonintegrable dynamical systems contains different areas with the topologically different characteristic phase trajectories. Paradigmatic model
of the complex, chaotic systems with the minimal chaos is the perturbed mathematical pendulum. The mathematical pendulum is exactly integrable in the absence of the time-dependent external driving.
However, when a time dependent perturbation is applied, dynamics in the vicinity of the separatrix become chaotic. For more details, one can refer to \cite{Zaslavsky}. \\

\section{Quantum parametrical resonance and Mathieu-Schr\"odinger equation}
Atom in the external electric field can be described via the driven nonlinear oscillator model (Lorentz's model):

\begin{equation}
H\big(x,p,t\big)=H_{0}\big(x,p\big)+H_{NL}+\varepsilon V\big(x,t\big),
\label{Lorentz}
\end{equation}

where
\begin{equation}
H_{0}=1/2\bigg(\frac{p^{2}}{m}+\omega_{0}^{2}mx^{2}\bigg),~~~H_{NL}=\beta x^{3}+\mu x^{4}+...,
\label{hamiltonian1}
\end{equation}

\begin{equation}
V\big(x,t\big)=V_{0}x\cos\Omega t,~~~\varepsilon V_{0}=ef_{0},~~\varepsilon \ll 1.
\label{hamiltonian2}
\end{equation}
Here $x$ and $p$ are the position and momentum of the electron, $\omega_{0}$ is the frequency of the oscillations, $\beta$ and $\mu$ are constants of the nonlinear terms.
We note that in the regime of moderate nonlinearity, in the nonlinear term $H_{NL}$  is enough to retain $\beta x^{3},~~\mu x^{4}$ terms only \cite{Zaslavsky}.
By means of the transformation to the canonical action--angle variables $x=\big(2I/m\omega_{0}\big)^{1/2}\cos(\theta)~~p=-\big(2Im\omega_{0}\big)^{1/2}\sin(\theta)$ and assuming that the resonance
condition holds $\Omega=\omega_{0}$ one can deduce the transformed Hamiltonian

\begin{equation}
H=H_{0}(I)+\varepsilon V(I)\cos(\varphi).
\label{transformd}
\end{equation}
where

\begin{equation}
H_{0}(I)=\omega_{0}I+H_{NL},~~~H_{NL}=\frac{3\pi}{4}\frac{I^{2}}{m\omega_{0}^{2}}\mu,
\label{2transformd}
\end{equation}
and
\begin{equation}
\varphi=\theta-\omega t, ~~~\varepsilon V(I)=V_{0}\sqrt{I/m\omega_{0}}.
\label{3transformd}
\end{equation}

We assume that deviation of the action $\Delta I=I-I_{0}$ from the nonlinear resonance condition 
$\omega_{0}+\omega_{NL}(I_{0})=\Omega$, $\omega_{NL}=(3\pi/2)(I\mu/m\omega_{0}^{2})$
is small. After implementing the series expansion finally we obtain

\begin{equation}
H=\frac{\omega'}{2}\big(\triangle I\big)^{2}+U\cos\varphi .
\label{4transformd}
\end{equation}

Here $\omega'=\big(d\omega_{NL}(I)/dI\big)\mid_{I=I_{0}}$, $U=\varepsilon V\big(I_{0}\big)$.\\

The classical phase space of the Hamiltonian (\ref{4transformd}) consists of the two topologically different domains:
Domains of the closed and open phase trajectories divided by area of separatrix. Thus the solution of the classical problem shows bifurcation tendency. Namely, the solution drastically depends
on the total energy of the system $E$ and in the explicit form read:

\begin{equation}
\Delta I=\sqrt{\big(E+U\big)\omega'}\dn\bigg(\omega'\sqrt{\big(E+U\big)\omega't},k\bigg), E>U
\label{classical1}
\end{equation}

\begin{equation}
\Delta I=\sqrt{\big(E+U\big)\omega'}\cn\bigg(\omega'\sqrt{\big(E+U\big)\omega't},1/k\bigg), E<U.
\label{classical2}
\end{equation}
Here $\dn(u,\varphi)$ and $\cn(u,\varphi)$ are the Jacobian delta amplitude and Jacobian elliptic cosine respectively.
Frequency of the system $\omega\big(I\big)=\pi/\ln\big(32/(1-E)\big)$  diverges logarithmically in the vicinity of the separatrix $k=\sqrt{2U/\big(E+U\big)}=1$.
Equilibrium points are defined via condition $p_{s}=0,~~~\frac{dU(q_{s})}{dq}=0$. In the vicinity of the equilibrium point $p-p_{s}=\pm \bigg(E-E_{s}-\frac{1}{2}\frac{d^{2}U(q_{s})}{dq^{2}}\big(q_{s}\big)\big(q-q_{s}\big)^{2}\bigg)^{1/2}$. Our particular interest concerns hyperbolic equilibrium points where motion is unstable $\frac{d^{2}U(q_{s})}{dq^{2}}<0$. When a time dependent perturbation is applied, in the vicinity of the separatrix appears stochastic layer and the complex homoclinic structure.
The width of the layer is proportional to the perturbation strength. \\

Due to the fundamental principle of the correspondence, one could expect to see the nontrivial behavior of the system in the quantum case as well.
Transition to the quantum case can be performed through the substitution $\Delta I\rightarrow -\imath \hbar\partial/\partial\varphi$
and after a little algebra we deduce Mathieu-Schr\"odinger equation:

\begin{equation}
\frac{d^{2}\psi_{n}}{d\varphi^{2}}+\big(E_{n}-V(l,\varphi)\big)\psi_{n}=0 .
\label{Mathieuequation}
\end{equation}
Here $V(l,\varphi)=2l\cos2\varphi$ and we rescaled energy, potential barrier and angle respectively: $E_{n}\rightarrow \frac{8E_{n}}{\hbar^{2}\omega'}$, $l\rightarrow \frac{8U}{\hbar^{2}\omega'}$, $\varphi\rightarrow 2\varphi$.\\

We note that Mathieu-Schr\"odinger equation can be derived in a less formal way by considering model of quantum nonlinear oscillator interacting with the strong electric field \cite{Berman}.
The detailed analysis of the Mathieu-Schr\"odinger equation (\ref{Mathieuequation}) was done in the references \cite{chotorlishvili1,chotorlishvili}. The energy spectrum
of the Mathieu-Schr\"odinger equation parametrically depends on the potential barrier $E_{n}\big(l\big)$ and contains two degenerate and one non-degenerate domain.
The main discovery of \cite{chotorlishvili} is the link between quantum parametric resonance and Klein's four-group. Namely transformation operations
\begin{equation}
\begin{split}
G\big(\varphi\rightarrow -\varphi\big)=a,~~~G\big(\varphi\rightarrow \pi-\varphi\big)=b, \\
G\big(\varphi\rightarrow \pi+\varphi\big)=c,~~~G\big(\varphi\rightarrow \varphi\big)=e,
\label{Mathieuequation2}
\end{split}
\end{equation}
of the Mathieu functions $\ce_{n}(\varphi),~~\se_{n}(\varphi)$ form the Klein's four-group $G$ with the following three invariant subgroups
\begin{equation}
\begin{split}
G_{0}\subset e,a, \\
G_{+}\subset e,b, \\
G_{+}\subset e,c. \\
\label{Klein}
\end{split}
\end{equation}

Irreducible representation basis functions of the subgroup $G_{0}$ formed by Mathieu functions correspond to the non-degenerate energy spectrum $\xi_{n}(\varphi,l)= \ce_{n}(l,\varphi),~~\se_{n}(l,\varphi)$,
while irreducible representation basis functions $\phi_{n}^{\pm}(\varphi,l)= \frac{\sqrt{2}}{2}\big(\ce_{n}(l,\varphi)\pm i \se_{n}(l,\varphi)\big)$ and $\psi_{n}^{\pm}(\varphi,l)= \frac{\sqrt{2}}{2}\big(\ce_{n}(l,\varphi)\pm i \se_{n+1}(l,\varphi)\big)$ of the two other subgroups $G_{-},G_{+}$ correspond to the degenerate energy spectrum. \\

\begin{figure}[h!]
\begin{center}
\includegraphics[width=0.99\linewidth]{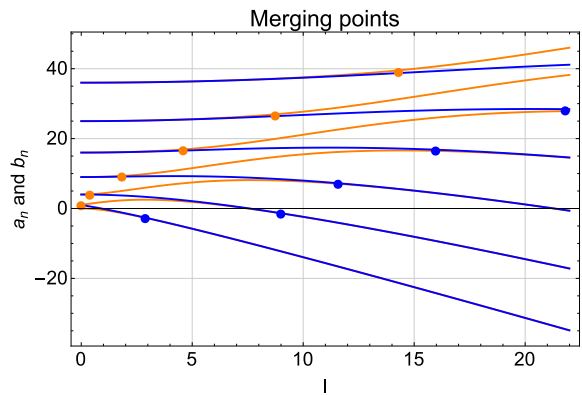}
\caption{Parametric dependence of the energy spectrum $E_{n}\big(l\big)$ of the Mathieu Scrhodinger equation (Mathieu characteristics) on the barrier height $l$.
Splitting and merging points define bounders of the $G_{-},G_{+}$ and $G_{0}$ subgroups.}
\label{Mathieucharacteristics}
\end{center}
\end{figure}

These three domains $G_{-},~G_{+}$ and $G_{0}$ on the parametric space $\big(E_{n}(l),l\big)$ are divided by splitting and merging points of the Mathieu characteristics (see
Fig. (\ref{Mathieucharacteristics})). The slight variation of barrier's height
$l(t)=l_{n}^{c}\pm \triangle l \cos(\omega t)$ in the vicinity of the splitting and merging points $l_{n}^{c}$ leads to the abrupt changes in the symmetry of the system.
The key issue is that the values $l=l_{n}^{c}+ \triangle l $ and $l=l_{n}^{c}- \triangle l$
belong to the different symmetry subgroups $G_{-},~G_{+},~G_{0}$. Note, that the
values of the splitting and merging points can be defined precisely for each quantum level (See TABLE I and TABLE II).

\begin{table}[h!]
\caption{Splitting points of the parameter $l$ for the transition from the $G_{-}$ subgroup to the  $G_{0}$ subgroup for different energy levels $E_{n}$.}
\begin{center}
$
\begin{array}{|l|c|c|c|c|c|c|c|c|}
\hline
n & 1 & 2 & 3 & 4 & 5 & 6 & 7 & 8 \\
\hline
l_c & 0.0 & 0.2 & 1.14 & 3.17 & 6.42 & 10.95 & 16.78 & 23.93 \\
\hline
\end{array}
$
\end{center}
\end{table}

\begin{table}[h!]
\caption{Merging points of the parameter $l$ for the transition from the $G_{0}$ subgroup to the  $G_{+}$ subgroup for different energy levels $E_{n}$.}
\begin{center}
$
\begin{array}{|l|c|c|c|c|c|c|c|c|}
\hline
n & 1 & 2 & 3 & 4 & 5 & 6 & 7 & 8 \\
\hline
l_c & 3.42 & 7.51 & 13.93 & 18.4 & 24.69 & 32.23 & 40.96 & 50.84 \\
\hline
\end{array}
$
\end{center}
\end{table}

\section{Switching of the symmetry subgroup and observable quantities}

Our primary interest concerns the question whether the switching of the symmetry subgroup may lead to the trembling motion. To answer this question we
explore expectation values of the commutator $v_{\varphi}=\dot{\varphi}=1/i \big[\varphi,H\big]$ that means:

\begin{equation}
\begin{split}
\big<\phi_{n}^{\pm} (\varphi,l_{n}^{c}-\Delta l)\mid\dot{\varphi}\mid\phi_{n}^{\pm}(\varphi,l_{n}^{c}-\Delta l)\big>, \\
\big<\xi_{n} (\varphi,l_{n}^{c}+\Delta l)\mid\dot{\varphi}\mid\xi_{n}(\varphi,l_{n}^{c}+\Delta l)\big>,
\label{commutator1}
\end{split}
\end{equation}

\begin{equation}
\begin{split}
\big<\xi_{n}(\varphi,l_{n}^{c}-\Delta l)\mid\dot{\varphi}\mid\xi_{n}(\varphi,l_{n}^{c}-\Delta l)\big>.\\
\big<\psi_{n}^{\pm}(\varphi,l_{n}^{c}+\Delta l)\mid\dot{\varphi}\mid\psi_{n}^{\pm}(\varphi,l_{n}^{c}+\Delta l)\big>.
\label{commutator2}
\end{split}
\end{equation}

Here $l_{n}^{c}$ in Eq. (\ref{commutator1})) are the splitting points corresponding to the transitions $G_{-}\rightarrow G_{0}$,
while merging points $l_{n}^{c}$ in Eq. (\ref{commutator2})) correspond to the transition $G_{0}\rightarrow G_{+}$. For short, the following notations of the irreducible basis functions are adopted:

\begin{equation}
G^- \to \phi^\pm_n (\varphi,l) = \frac{\sqrt{2}}{2}\big(\ce_n(l,\phi) \pm i \se_n(l,\varphi)\big),
\end{equation}

\begin{equation}
G^0 \to \xi_n(\varphi,l) =  \ce_n(l,\varphi), \mbox{ or }  \eta_n(\varphi,l) = \se_n(l,\varphi),
\end{equation}

and

\begin{equation}
G^+ \to \psi^\pm_n (\varphi,l) = \frac{\sqrt{2}}{2}\big(\ce_n(l,\varphi) \pm i \se_{n+1}(l,\varphi)\big).
\end{equation}

In order to explore the effect of symmetry switching,
we evaluate Eq. (\ref{commutator1})) and Eq. (\ref{commutator2})) in the limit of $\Delta l\rightarrow 0$.

The expressions for the expectation value of the velocity operator can be further simplified using trigonometric representation of Mathieu functions\cite{bateman}:

\begin{equation}
\ce_{2m}(l,\phi) = \sum_{r=0}^\infty A_{2r}^{(2m)}(l) \cos(2r\varphi),
\label{furier1}
\end{equation}

\begin{equation}
\ce_{2m+1}(l,\phi) = \sum_{r=0}^\infty A_{2r+1}^{(2m+1)}(l) \cos((2r+1)\varphi),
\label{furier2}
\end{equation}

\begin{equation}
\se_{2m+1}(l,\phi) = \sum_{r=0}^\infty B_{2r+1}^{(2m+1)}(l) \sin((2r+1)\varphi),
\label{furier3}
\end{equation}

\begin{equation}
\se_{2m+2}(l,\phi) = \sum_{r=0}^\infty B_{2r+2}^{(2m+2)}(l) \sin((2r+2)\varphi).
\label{furier4}
\end{equation}

Here  $A_{2r}^{(2m)}(l)$, $A_{2r+1}^{(2m+1)}(l)$ and $B_{2r+1}^{(2m+1)}(l)$, $B_{2r+2}^{(2m+2)}(l)$
are the Fourier coefficients that depend on the quantum number $m$ and the barrier height $l$.
We note that the trembling should occur directly at the bifurcation (splitting/merging) points.
We are interested in the estimation of the velocity increment in the bifurcation point $\Delta v_{\varphi}=v_{\varphi}(G_{0},l= l_{c}+\triangle l)_{\triangle l\rightarrow 0}-v_{\varphi}(G_{-},l= l_{c}-\triangle l)_{\triangle l\rightarrow 0}$, where
$v_{\varphi}(G_{-},l= l_{c}-\triangle l)_{\triangle l\rightarrow 0}$ is the expectation value of velocity before passing the bifurcation point and $v_{\varphi}(G_{0},l= l_{c}+\triangle l)_{\triangle l\rightarrow 0}$
is the expectation value of velocity in the subgroup $G_{0}$ after passing the bifurcation point. \\

Taking into account (\ref{furier1})-(\ref{furier4}) we derive analytical expressions of the
expectation values of velocity operator at the point $l_c$ corresponding to the switching of the symmetry subgroups.
In particular, for the symmetry switching $G^-\to G^0$, in the limit $l \to l_c$ we deduce that:

\begin{itemize}
  \item
  for the states $\phi_{2n+1}^\pm(l,\varphi)$

\begin{multline}
\expval{\hat{v}}_{l \to l_c^+} = -2i \times\\
  \bra{\phi^\pm_{2n+1}(l_c-\Delta l,\varphi)} \dvarphi \ket{\phi^\pm_{2n+1}(l_c-\Delta l,\varphi)}_{\Delta l \to 0}
  \\
  =  \mp 4 \sum_{r=0}^\infty (2r+1) A_{2r+1}^{(2n+1)}(l_c) B_{2r+1}^{(2n+1)}(l_c),
\label{velocity}
\end{multline}

\item
for the states $\xi_{2n+1}(l,\varphi)$
\begin{multline}
  \expval{\hat{v}}_{l \to l_c^-} = -2i \times\\
    \bra{\xi_{2n+1}(l_c+\Delta l,\varphi)} \dvarphi \ket{\xi_{2n+1}(l_c+\Delta l,\varphi)}_{\Delta l \to 0}
    \\
    = 0,
\end{multline}

\item
for the states $\eta_{2n+1}(l,\varphi)$
\begin{multline}
  \expval{\hat{v}}_{l \to l_c^+} = -2i \times\\
    \bra{\eta_{2n+1}(l_c+\Delta l,\varphi)} \dvarphi \ket{\eta_{2n+1}(l_c+\Delta l,\varphi)}_{\Delta l \to 0}
    \\
    = 0.
\end{multline}

\end{itemize}

The above expressions allow us to determine the jumps in the value of velocity at the bifurcation point for different transitions between $G_{-}$ and $G_{0}$ states.
Results of calculations are presented in Table~\ref{tableDELTAV}.  We note that trembling occurs only because of the fact that separatrix line is the border between different symmetry subgroups of the Mathieu-Schr\"odinger equation. Eigenfunctions are smooth functions of the barrier height and therefore within the subgroups effect of the trembling is absent.

The results for the quantum states $\Psi_{2n+1}^\pm(l,\varphi)$, $\Psi_{2n,2}^\pm(l,\varphi)$ can be obtained in the similar way (not shown for shortness).
An interesting fact is the absence of the trembling during transitions between subgroups $G_{0}\rightarrow G_{+}$. The reason is quite clear. Transition $G_{0}\rightarrow G_{+}$ occurs in the  limit of high potential barrier and this naturally suppress kinetic effects. \\

To infer increment in the squared velocity that occurs in the bifurcation point, we calculate expectation value of the squared velocity operator $\big(\Delta v^{2}_{\varphi}\big)_{n}$. In particular we estimate the jump that occurs due to the symmetry switching  $\Delta v^{2}_{\varphi}=v^{2}_{\varphi}(G_{-}\mapsto G_{0})-v^{2}_{\varphi}(G_{-})$. After straightforward calculations in the limit $l \to l_c$ we deduce:
\begin{itemize}
  \item
  for the states $\phi_{2n+1}^\pm(l,\varphi)$
\begin{multline}
\expval{\hat{v}^2}_{l \to l_c^+} = -4 \times\\
  \bra{\phi^\pm_{2n+1}(l_c-\Delta l,\varphi)} \d2varphi \ket{\phi^\pm_{2n+1}(l_c-\Delta l,\varphi)}_{\Delta l \to 0}
  \\
  =  \mp 8 \sum_{r=0}^\infty (2r+1)^2
    \left(
      \left(A_{2r+1}^{(2n+1)}(l_c)\right)^2 +
      \left(B_{2r+1}^{(2n+1)}(l_c)\right)^2
    \right),
\label{v2phi}
\end{multline}

  \item
  for the states $\xi_{2n+1}(l,\varphi)$
\begin{multline}
\expval{\hat{v}^2}_{l \to l_c^+} = -4 \times\\
  \bra{\xi_{2n+1}(l_c+\Delta l,\varphi)} \d2varphi \ket{\xi_{2n+1}(l_c+\Delta l,\varphi)}_{\Delta l \to 0}
  \\
  =  \mp 4 \sum_{r=0}^\infty (2r+1)^2 \left(A_{2r+1}^{(2n+1)}(l_c)\right)^2.
\label{v2xi}
\end{multline}

\item
for the states $\eta_{2n+1}(l,\varphi)$
\begin{multline}
\expval{\hat{v}^2}_{l \to l_c^+} = -4 \times\\
\bra{\eta_{2n+1}(l_c+\Delta l,\varphi)} \d2varphi \ket{\eta_{2n+1}(l_c+\Delta l,\varphi)}_{\Delta l \to 0}
\\
=  \mp 4 \sum_{r=0}^\infty (2r+1)^2 \left(A_{2r+1}^{(2n+1)}(l_c)\right)^2.
\label{v2eta}
\end{multline}

\end{itemize}

Result of calculation for the jumps in the expectation values of the squared velocity at the bifurcation point for different transitions between symmetry subgroups $G_{-}$ and $G_{0}$
are summarized in Table~\ref{tableDELTASQUAREV}.

\begin{figure}[h!]
\begin{center}
\includegraphics[width=0.99\linewidth]{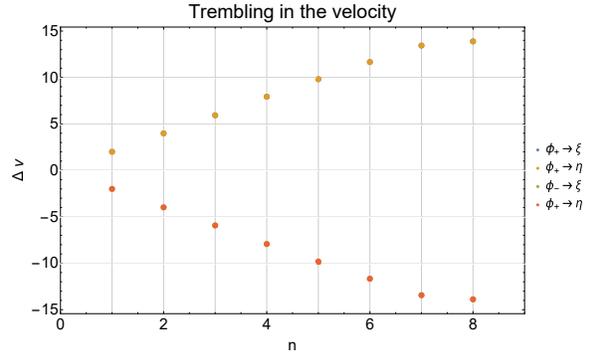}\\
\caption{Increment in velocity for the transition between symmetry subgroups
$G_{-}\rightarrow G_{0}$  between the states
$\phi^\pm_{n}(l,\varphi)\rightarrow \xi_{n}(l,\varphi)$ (blue dots) and
states $\phi^\pm_{n}(l,\varphi)\rightarrow \eta_{n}(l,\varphi)$ (yellow dots).}
\label{Velocity increment}
\end{center}
\end{figure}

It is easy to see that the mean increment of the squared velocity
(Eqs.(\ref{v2phi})-(\ref{v2eta}))
$\Delta v^{2}_{\varphi}$ is not equal to the square of the increment of mean velocity
(Eq.(\ref{velocity})) $\big(\Delta v_{\varphi}\big)^{2}$.
Taking into account Eq.(\ref{velocity})-(\ref{v2eta}) we calculate mean square
fluctuations $F=\sqrt{\Delta v^{2}_{\varphi}-\big(\Delta v_{\varphi}\big)^{2}}$ of
the velocity increment (see TABLE III and TABLE IV). Apparently, fluctuations increase
with the quantum number $n$ (see Fig. (\ref{Velocity increment})). Our results confirm that not only
free particle but particle trapped in the potential well being in the excited quantum
states can experience trembling motion. \\

\begin{table}[h!]
\begin{center}
\caption{Jump in the velocity for the transitions $G_{-}\rightarrow G_{0}$ between the states $\phi^\pm_{n}(l,\varphi)\rightarrow \ce_{n}(l,\varphi)$.}
$$
\begin{array}{|c|cccc|}
\hline
    & \multicolumn{4}{c|}{(\Delta v)_{l\to l_c}}  \\
    \cline{2-5}
  n & \phi_{+} \to \xi & \phi_{+} \to \eta & \phi_{-} \to \xi & \phi_{-} \to \eta \\
\hline
\hline
1 & 2. & 2. & -2. & -2. \\
2 & 3.981 & 3.981 & -3.981 & -3.981 \\
3 & 5.929 & 5.929 & -5.929 & -5.929 \\
4 & 7.927 & 7.927 & -7.927 & -7.927 \\
5 & 9.815 & 9.815 & -9.815 & -9.815 \\
6 & 11.665 & 11.665 & -11.665 & -11.665 \\
7 & 13.437 & 13.437 & -13.437 & -13.437 \\
8 & 13.884 & 13.884 & -13.884 & -13.884 \\
 \hline
 \end{array}
 $$
 \label{tableDELTAV}
 \end{center}
 \end{table}

 \begin{table}[h!]
 \begin{center}
 \caption{Jump in the square of the velocity for the transitions $G_{-}\rightarrow G_{0}$ between the states $\phi^\pm_{n}(l,\varphi)\rightarrow \ce_{n}(l,\varphi)$.}
 $$
 \begin{array}{|c|cccc|}
 \hline
     & \multicolumn{4}{c|}{(\Delta v^2 )_{l\to l_c}}  \\
     \cline{2-5}
   n & \phi_{+} \to \xi & \phi_{+} \to \eta & \phi_{-} \to \xi & \phi_{-} \to \eta \\
 \hline
 \hline
1 &  0. & 0. & 0. & 0. \\
2 &  0.088 & -0.087 & 0.088 & -0.087 \\
3 &  0.204 & -0.203 & 0.204 & -0.203 \\
4 &  0.079 & -0.08 & 0.079 & -0.08 \\
5 &  0.181 & -0.182 & 0.181 & -0.182 \\
6 &  0.304 & -0.304 & 0.304 & -0.304 \\
7 &  0.558 & -0.558 & 0.558 & -0.558 \\
8 &  10.964 & -10.964 & 10.964 & -10.964 \\
   \hline
  \end{array}
  $$
  \label{tableDELTASQUAREV}
  \end{center}
  \end{table}

\begin{figure}[h!]
\begin{center}
\includegraphics[width=0.99\linewidth]{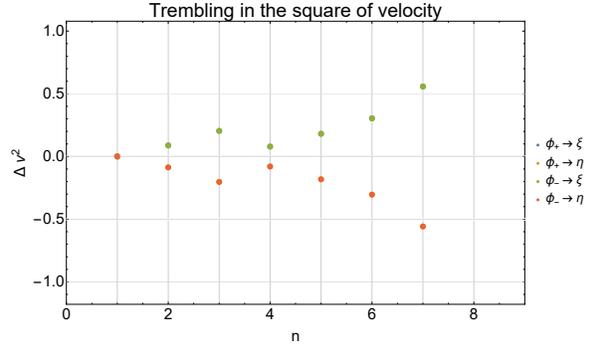}\\
\caption{Jump in the square of velocity for the transition between symmetry subgroups
$G_{-}\rightarrow G_{0}$ and particular between the states
$\phi^\pm_{n}(l,\varphi)\rightarrow \xi_{n}(l,\varphi)$ (blue dots) and
states $\phi^\pm_{n}(l,\varphi)\rightarrow \eta_{n}(l,\varphi)$ (yellow dots).}
\label{}
\end{center}
\end{figure}

\section{Uncertainty relations and crossover with trembling motion}

Overwhelmingly under "uncertainty" of arbitrary quantity $\hat{A}$ mean square deviation is meant $<\hat{A}^{2}>-<\hat{A}>^{2}$.
This formulation was established by the works of Heisenberg. In the present work, we are interested in the question whether there exist crossover between quenching of symmetry
and principle of uncertainty? Note that the derivation of the uncertainty relations implicitly implies that the self-adjoint operators are defined on the same set of the basis function.
For angular momentum operator and an angular variable, this is not the case. Operator adjoint to the angular momentum operator should be a periodical function of the angular variable.
This problem is precisely studied in the literature, see \cite{carruthers} and references therein. Here we will follow a formalism described in \cite{carruthers} in details and
try to find the crossover between uncertainty relations and trembling in the context of the quenching symmetry. On an intuitive level, albeit this crossover is predictable. However,
our promise is to provide more rigorous arguments in support. Uncertainty relations between z-component of the angular momentum operator $L_{z}=-i\frac{\partial}{\partial\varphi}$ and angular variable
$\varphi$ can be quantified as follows \cite{carruthers}

\begin{figure}[h!]
\begin{center}
\includegraphics[width=0.99\linewidth]{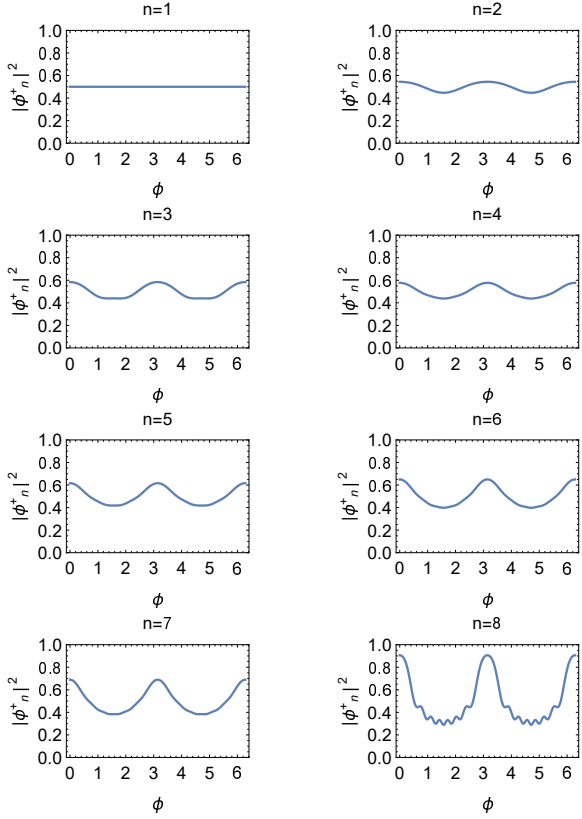}\\
\caption{Probability density for $\phi^+_{n}(l,\varphi)$, $n = 1, 2, ..., 8$.}
\label{wavefunctions}
\end{center}
\end{figure}

\begin{equation}
ur_a = (\Delta L_z )^2 (\Delta \sin\varphi)^2 - \frac{1}{4}(\Delta \cos\varphi)^2 \ge 0,
\label{Uncertainty1}
\end{equation}
and
\begin{equation}
ur_b = (\Delta L_z )^2 (\Delta \cos\varphi)^2 - \frac{1}{4}(\Delta \sin\varphi)^2 \ge 0.
\label{Uncertainty2}
\end{equation}
Here we introduced the following notations:
\begin{multline}
(\Delta \sin\varphi)^2 = \expval{\sin^2\varphi} - \expval{\sin\varphi}^2,
\end{multline}
\begin{multline}
(\Delta \cos\varphi)^2 = \expval{\cos^2\varphi} - \expval{\cos\varphi}^2,
\end{multline}
\begin{multline}
(\Delta \hat{L}_z)^2 = \expval{\hat{L}_z^2} - \expval{\hat{L}_z}^2.
\end{multline}

\begin{table}[ht!]
\begin{center}
\caption{Values of the expression $ur_a$ for
$\phi_{+}$, $\phi_{-}$ (region $G_{-}$), $\xi$ and $\eta$ (region $G_{0}$).}
$$
\begin{array}{|c|cccc|}
\hline
\text{n} & \phi _+ & \phi _- & \xi  & \eta  \\
\hline
1 & -0.125 & -0.125 & 0.0625 & 0.6875 \\
2 & -0.119738 & -0.119738 & 1.59984 & 1.92984 \\
3 & -0.0795267 & -0.0795267 & 3.81139 & 4.15142 \\
4 & -0.0617743 & -0.0617743 & 7.20129 & 7.31649 \\
5 & 0.0777876 & 0.0777876 & 10.8688 & 11.0587 \\
6 & 0.303495 & 0.303495 & 15.1676 & 15.4109 \\
7 & 0.68743 & 0.68743 & 19.8053 & 20.2087 \\
8 & 2.69405 & 2.69405 & 17.8697 & 23.1462 \\
\hline
\end{array}
$$
\label{tableURA}
\end{center}
\end{table}

\begin{table}[ht!]
\begin{center}
\caption{Values of the expression $ur_b$ for
$\phi_{+}$, $\phi_{-}$ (region $G_{-}$), $\xi$ and $\eta$ (region $G_{0}$).}
$$
\begin{array}{|c|cccc|}
\hline
\text{n} & \phi _+ & \phi _- & \xi  & \eta  \\
\hline
1 & -0.125 & -0.125 & 0.6875 & 0.0625 \\
2 & -0.106602 & -0.106602 & 2.11391 & 1.82766 \\
3 & -0.0514835 & -0.0514835 & 4.79486 & 4.55658 \\
4 & -0.0343079 & -0.0343079 & 8.39221 & 8.31676 \\
5 & 0.147406 & 0.147406 & 13.3947 & 13.2956 \\
6 & 0.454449 & 0.454449 & 19.5324 & 19.4411 \\
7 & 1.00283 & 1.00283 & 26.8837 & 26.7593 \\
8 & 4.80609 & 4.80609 & 35.0808 & 35.2863 \\
\hline
\end{array}
$$
\label{tableURB}
\end{center}
\end{table}

Taking into account that $\hat{L}_z = -i \frac{\partial}{\partial\varphi} = \frac{1}{2} \hat{v}_\varphi$,
$\hat{L}^2_z = \frac{1}{4} \hat{v^2}_\varphi$ and using Eqs.~(\ref{furier1})-(\ref{furier4})
it is easy to determine the analytical expressions for expectation values of the operators required
to calculate $ur_a$ and $ur_b$
\begin{multline}
\bra{\phi^\pm_{n}(l,\varphi)} \sin\varphi \ket{\phi^\pm_{n}(l,\varphi)}_{l\to l_c^+} =  0,
\end{multline}
\begin{multline}
\bra{\phi^\pm_{n}(l,\varphi)} \cos\varphi \ket{\phi^\pm_{n}(l,\varphi)}_{l\to l_c^+} =  0,
\end{multline}

\begin{multline}
\bra{\xi_{n}(l,\varphi)} \sin\varphi \ket{\xi_{n}(l,\varphi)}_{l\to l_c^-} =  0,
\end{multline}
\begin{multline}
\bra{\eta_{n}(l,\varphi)} \sin\varphi \ket{\eta_{n}(l,\varphi)}_{l\to l_c^-} =  0,
\end{multline}
\begin{multline}
\bra{\xi_{n}(l,\varphi)} \cos\varphi \ket{\xi_{n}(l,\varphi)}_{l\to l_c^-} =  0,
\end{multline}
\begin{multline}
\bra{\eta_{n}(l,\varphi)} \cos\varphi \ket{\eta_{n}(l,\varphi)}_{l\to l_c^-} =  0,
\end{multline}

\begin{multline}
\bra{\xi_{n}(l,\varphi)} \sin^2\varphi \ket{\xi_{n}(l,\varphi)}_{l\to l_c^-} =
\frac{1}{2}\sum_{r=0}^{\infty} (A_{2r+1})^2,
\end{multline}

\begin{multline}
\bra{\eta_{n}(l,\varphi)} \sin^2\varphi \ket{\eta_{n}(l,\varphi)}_{l\to l_c^-} =
\frac{1}{2}\sum_{r=0}^{\infty} (B_{2r+1})^2,
\end{multline}

\begin{multline}
\bra{\xi_{n}(l,\varphi)} \cos^2\varphi \ket{\xi_{n}(l,\varphi)}_{l\to l_c^-} =
\frac{1}{2}\sum_{r=0}^{\infty} (A_{2r+1})^2,
\end{multline}

\begin{multline}
\bra{\eta_{n}(l,\varphi)} \cos^2\varphi \ket{\eta_{n}(l,\varphi)}_{l\to l_c^-} =
\frac{1}{2}\sum_{r=0}^{\infty} (B_{2r+1})^2,
\end{multline}

\begin{multline}
\bra{\phi^\pm_{n}(l,\varphi)} \sin^2\varphi \ket{\phi^\pm_{n}(l,\varphi)}_{l\to l_c^+} =\\
\frac{1}{4}\sum_{r=0}^{\infty} (A_{2r+1})^2 + (B_{2r+1})^2,
\end{multline}

\begin{multline}
\bra{\phi^\pm_{n}(l,\varphi)} \cos^2\varphi \ket{\phi^\pm_{n}(l,\varphi)}_{l\to l_c^+} =\\
\frac{1}{4}\sum_{r=0}^{\infty} (A_{2r+1})^2 + (B_{2r+1})^2,
\end{multline}

Further simplification of Eq. (\ref{Uncertainty1}), Eq. (\ref{Uncertainty2}) relies on the fact that wave functions have a certain maximum for particular values of the angle $\varphi = n\pi$, n=0,1,2,..N. This fact mimics the quantum counterpart of the classical dynamical systems. Namely, during motion, particle spends the major part of time in the vicinity of the hyperbolic equilibrium points.
As we see this effect is more profound for the high exited states and is absent in the ground state (see Fig. (\ref{wavefunctions})). Thus, when studying uncertainty relations, we are interested
in the excited states. After expending Eq.(\ref{Uncertainty1}) and Eq.(\ref{Uncertainty2}) in the vicinity of the maximum points, we immediately see that the second equation holds automatically while the first reduced to the
following form:

\begin{equation}
\big(\Delta L_{z}\big)\big<\varphi^{2}\big>\geq \frac{1}{4}\big(1-\frac{1}{2}\big<\varphi^{2}\big>\big).
\end{equation}

The values of the expression $ur_a$ and $ur_b$ for particular states and transitions
between subgroups $G_{-}$ and $G_{0}$ are tabulated in Tables~\ref{tableURA} and~\ref{tableURB}.
As we see that switching of the symmetry subgroup leads not only to the Zitterbewegung but also accompanied by dramatic changes in the uncertainty relations. In particular
we see jump in the expectation values (32) and (33). \\

\section{Torsional oscillation in polyatomic molecule}

Results obtained in the previous sections have a certain physical application in polyatomic molecules.
It is well-known that polyatomic molecules can perform an internal rotational motion of two types: torsional oscillation and free rotation of one part of the molecule with respect to the other part \cite{Flyger}.
On the phase plane, these two types of motions are separated by the separatrix line. Hamiltonian of the polyatomic molecule related to the internal rotation has the following form \cite{Flyger}:

\begin{equation}
\hat{H}=-\frac{\hbar^{2}}{2I}\frac{d^{2}}{d\varphi^{2}}+U\big(\varphi\big).
\label{torsion1}
\end{equation}

Here $I=I_{1}I_{2}/\big(I_{1}+I_{1}\big)$ is the reduced moment of inertia, $I_{1},~I_{2}$ are the moments of inertia of the rotating parts of the molecule with respect to symmetry axis.
Potential energy has the form:
\begin{equation}
U\big(\varphi\big)=\frac{V_{0}}{2}\big(1-\cos(n\varphi)\big).
\label{torsion2}
\end{equation}
Here $V_{0}$ defines the height of potential barrier that separates the torsional oscillations from the rotation of one part of the molecule with respect to the other part,
and $n$ defines the quantity of equilibrium orientations of one part of the molecule with respect to the other part \cite{Flyger}.
Paradigmatic model of organic molecules characterized by the property of internal rotation is the molecule of ethane $C_{2}H_{6}$ with the corresponding
parameters: $I_{1}=I_{2}=5.3\cdot 10^{-47}$kg\,m$^{2}$, $V_{0}=2.1\cdot 10^{-20}$J.
Using transformations $\varphi\mapsto n\varphi/2$,  $E\mapsto\frac{8I}{n^{2}\hbar^{2}}\big(E-V_{0}/2\big)$
Eq.~(\ref{torsion1}) can be easily mapped into the Mathieu-Schr\"odinger equation Eq(\ref{Mathieuequation}).
For the experiment we propose to use the monochromatic pumping field with the frequency $\Omega \ll V_{0}/\hbar$. Such a pumping field can cause a slow modulation of swift electron motion in a molecule.
The formation of an energy barrier is a result of averaging over swift electron motion and due to the pumping affect the value of barrier becomes time-dependent
$V(t)=V_{0}+\triangle V\cos\Omega t$.\\

\section{Conclusions}
The aim of the present work is to generalize the concept of Zitterbewegung for the systems that possess complex symmetry
properties. In particular, we discovered that trembling motion might occur due to the dramatic changes in the symmetry of the system. For this purpose, we exploit the paradigmatic model of quantum chaos, quantum mathematical pendulum. The symmetry group of this system is the Klein's four-group with three invariant subgroups. The energy spectrum of the system parametrically depends on the height of the potential barrier and contains degenerate and non-degenerate areas corresponding to the different symmetry subgroups. We observed that the changes in the potential barrier height switches the symmetry subgroup and lead to the trembling motion. We have shown that the trembling enhances for higher excited states which in turn proved by the analysis of the mean square fluctuations of the velocity operator. 

\section*{Acknowledgements}
This work was supported by the German Science Foundation, DFG under SFB 762. Two of us,
P.Z. and I.T., also acknowledge the support received from
Centre for Innovation and Engineering Knowledge at the University of Rzesz\'ow.


\newpage

\begin{thebibliography}{99}

\bibitem{Schrodinger} E. Schr\"odinger, Sitzungsb. Preuss. Akad. Wiss. Phys. Math. Kl. \textbf{24}, 418 (1930).
\bibitem{zawadzki20111} W. Zawadzki, T.M. Rusin, J. Phys. Condens. Matter \textbf{23} 143201 (2011).
\bibitem{Schliemann} J. Schliemann, D. Loss, and R. M. Westervelt   Phys. Rev. Lett. \textbf{94}, 206801 (2005).
\bibitem{Solano} L. Lamata, J. Leon, T. Sch\"atz, and E. Solano Phys. Rev.Lett. \textbf{98}, 253005 (2007).
\bibitem{Katsnelson} M. I. Katsnelson, K. S. Novoselov, A. K. Geim Nature Phys. \textbf{2}, 620 (2006).
\bibitem{Blatt} R. Gerritsma, G. Kirchmair, F. Z\"ahringer, E. Solano, R. Blatt, C. F. Roos Nature \textbf{463}, 68 (2010).
\bibitem{Engels} C. Qu, Chris Hamner, M. Gong, C. Zhang, and Peter Engels  Phys. Rev. A \textbf{88}, 021604(R) (2013).
\bibitem{LeBlanc} L. J. LeBlanc, M. C. Beeler, K. Jimenez-Garcya, A. R. Perry, S. Sugawa, R. A. Williams, and I. B. Spielman New Journal of Physics \textbf{15} 073011 (2013).
\bibitem{zawadzki2007} T. M. Rusin, and W. Zawadzki Phys. Rev. B \textbf{76}, 195439 (2007).
\bibitem{zawadzki2005} W. Zawadzki Phys. Rev. B \textbf{72}, 085217 (2005).
\bibitem{Portnoi} C. A. Downing and M. E. Portnoi Phys. Rev. A \textbf{90}, 052116 (2014).
\bibitem{Zaslavsky} G. Zaslavsky The Physics of Chaos in Hamiltonian Systems, Imperial College Press; 2 edition (2007).
\bibitem{Berman} G. P. Berman and G. M. Zaslavsky, Phys. Lett. \textbf{61} A, 295 (1977);
G. P. Berman, G. M. Zaslavsky, and A. R. Kolovsky, Phys. Lett. \textbf{87} A, 152 (1982).
\bibitem{carruthers} P. Carruthers, M.M. Nieto, Rev. Mod. Physics \textbf{40}, 411 (1968)
\bibitem{chotorlishvili1} A. Ugulava, L. Chotorlishvili, K. Nickoladze Phys. Rev. E \textbf{70}, 026219 (2004).
\bibitem{chotorlishvili} L. Chotorlishvili, A. Ugulava Physica D \textbf{239}, 103 (2010);
A. Ugulava, Z. Toklikishvili, S. Chkhaidze, R. Abramishvili, and L. Chotorlishvili Journal of Mathematical Physics \textbf{53}, 062101 (2012).
\bibitem{bateman} H. Bateman, Higher Transcendental Function, Vol. 3, Mc Graw-Hill (1955), p. 155.
\bibitem{Flyger} W.H.Flyger, Molecular Structure and Dynamics. New-Jersey, (1978);
T.Shimanouchi, Tables of Molecular Vibrational Frequencies Consolidated, National Bureau of Standards, \textbf{1}, 1-160, (1972).


\end{thebibliography}
\end{document}